\documentclass[a4paper]{jpconf}
\usepackage{graphicx}
\usepackage[reqno]{amsmath}
\usepackage{amssymb}

\def\ltap{\ \raisebox{-.4ex}{\rlap{$\sim$}} \raisebox{.4ex}{$<$}\ }

\newcommand{\bea}{\begin{equation}\begin{array}{c}}
\newcommand{\eea}{\end{array}\end{equation}}
\newcommand{\ea}{\end{array}} 
\newcommand{\beq}{\begin{equation}}
\newcommand{\eeq}{\end{equation}}
\newcommand{\bad}{\begin{array}{ccc}}

\newcommand{\ba}{\begin{array}{c}}

\newcommand{\half}{\frac{1}{2}}
\newcommand{\diag}{{\rm diag}}
\newcommand{\PMNS}{{\rm PMNS}}

\newcommand{\betabeta}{\mbox{$(\beta \beta)_{0 \nu}  $}}

\begin{document}
\title{Lepton Number Violation in TeV Scale See-Saw Extensions of the Standard Model}

\author{A~Ibarra$^{1}$, E~Molinaro$^{2,}$\footnote[5]{Talk given by the author at DISCRETE 2010, 6 December 2010, Rome, Italy.} and
S~T~Petcov$^{3,4,}$\footnote[6]{Also at: Institute of Nuclear Research and Nuclear Energy, 
Bulgarian Academy  of Sciences 1784 Sofia, Bulgaria.}}

\address{$^{1}$ Physik-Department T30d, Technische Universit\"at M\"unchen, James-Franck-Stra{\ss}e, 85748 Garching, Germany}
\address{$^{2}$ Centro de F\'{i}sica Te\'{o}rica de Part\'{i}culas (CFTP), Instituto Superior T\'{e}cnico, Avenida Rovisco Pais 1, 1049-001, Lisboa, Portugal}
\address{$^{3}$ SISSA and INFN-Sezione di Trieste, Via Bonomea 265, 34136 Trieste, Italy}
\address{$^{4}$ Institute for the Physics and Mathematics of the Universe, University of Tokyo, Kashiwa, 277-8583, Japan}

\begin{abstract}

The low-energy neutrino physics constraints 
on the TeV scale type I see-saw scenarios of neutrino mass generation are revisited.
It is shown that 
lepton charge ($L$) violation, associated to the production and decays
of heavy Majorana neutrinos $N_{j}$ having masses 
in the range of $M_j \sim (100 \div 1000)$ GeV and present 
in such scenarios, is hardly to be observed at ongoing and future particle accelerator 
experiments, LHC included, because of very strong constraints 
on the parameters and couplings responsible for the 
corresponding $|\Delta L| = 2$ processes.
If the heavy Majorana neutrinos $N_j$
are observed and they are associated only 
with the type I mechanism, 
they will behave effectively like pseudo-Dirac fermions.
Conversely, the observation of effects proving 
the Majorana nature of $N_j$ would imply that 
these heavy neutrinos
have additional relatively strong  
couplings to the Standard Model particles   
or that light neutrino masses compatible with the 
observations are generated 
by a mechanism other than see-saw 
(e.g., radiatively at one or two loop level) 
in which the heavy Majorana 
neutrinos $N_j$ are nevertheless involved. 
\end{abstract}

\section{Introduction}

Neutrino oscillation experiments~\cite{exper}~have 
revealed undeniable evidence for flavour non-conservation in the 
lepton sector. Flavour neutrino eigenstates, $\nu_{\ell L}$ ($\ell=e,\mu,\tau$),
oscillate, changing their lepton flavour quantum number \cite{BPont57,MNS62},
due to non-zero neutrino masses and neutrino mixing. The latter is 
parametrized by the well known Pontecorvo-Maki-Nakagawa-Sakata (PMNS) neutrino
mixing matrix \cite{Znu}. 

All present  neutrino oscillation data
can be described assuming 
3-flavour neutrino mixing in vacuum.
The number of massive neutrinos can, in general, be bigger than 3, 
if, for instance, there exist right-handed (RH)
sterile neutrinos \cite{BPont67} 
and they mix with the left-handed (LH) flavour neutrinos. 
It follows from the existing data that 
at least 3 of the neutrinos must be light, with masses
$m_1\neq m_2 \neq m_3$, $m_{1,2,3} \ltap 1$ eV and 
at least two of the $m_{j}$ different from zero.
At present there are no compelling 
experimental evidences for the existence  
of more than 3 light neutrinos. 

The smallness of neutrino masses, $m_{j}/m_{e}\lesssim 10^{-6}$, where
$m_{e}$ is the electron mass, suggests the 
presence of a new fundamental scale in 
particle physics and thus to new physics beyond the 
Standard Model (SM). Among the possible extensions of the SM
which can explain neutrino mass generation, a well known proposal is
the see-saw mechanism \cite{seesaw}. In its simplest formulation,
the type I see-saw scenario, at least two ``heavy'' $SU(2)_L\times U(1)_{Y}$ 
singlet RH neutrinos $\nu_{a R}$ are introduced in the theory.
Within the see-saw framework, the latter are 
assumed to possess a Majorana mass term as well as 
Yukawa type couplings with the Standard Model 
lepton and Higgs doublets. 

Below the electroweak (EW)
symmetry breaking scale, the light neutrino masses arise from the Lagrangian:
\begin{equation}
\mathcal{L}_{\nu}\;=\; -\, \overline{\nu_{\ell L}}\,(M_{D})_{\ell a}\, \nu_{aR} - 
\half\, \overline{\nu^{C}_{aL}}\,(M_{N})_{ab}\,\nu_{bR}\;+\;{\rm h.c.}\,, 
\label{typeI}
\end{equation}
%
where $\nu^{C}_{aL}\equiv C \overline{\nu_{aR}}^T$ ($a=1,\ldots,k$), $C$ being the charge 
conjugation matrix, $M_{N} = (M_{N})^T$ is the 
$k\times k$ Majorana mass matrix of the RH neutrinos,
and $M_{D}$ is the $3\times k$ neutrino Dirac 
mass matrix, which is generated by the matrix of 
neutrino Yukawa couplings after the  
EW symmetry breaking:   $M^{D} = v\lambda$, $\lambda_{\ell a}$ being 
the matrix of neutrino Yukawa couplings 
and $v = 174$ GeV being the v.e.v. of the SM Higgs doublet.
The matrices $M_{N}$ and 
$M_{D}$ are complex, in general.
A Majorana mass term $m_{\nu}$ for 
the active left-handed neutrinos is indeed generated 
by the interplay between the
Dirac mass term and the Majorana mass term 
of the heavy Majorana neutrinos. The well know see-saw mass relation is in this case:
$m_{\nu}\cong -  M_{D} M_{N}^{-1} (M_{D})^T$.

In grand unified theories, $M_{D}$ is typically of the
order of the charged fermion masses. In $SO(10)$ theories, 
for instance, $M_{D}$ coincides with the up-quark mass matrix.
Taking indicatively $m_{\nu} \sim 0.05$ eV, 
$M_{D}\sim 100$ GeV, one finds 
$M_{N}\sim 2\times 10^{14}$ GeV, which is close 
to the scale of unification of the electroweak and
strong interactions, $M_{GUT}\cong 2\times 10^{16}$ GeV.   
In GUT theories with RH neutrinos one finds 
that indeed the heavy Majorana neutrinos 
naturally obtain masses which are  
by few to several orders of magnitude smaller 
than $M_{GUT}$.   The estimate of $M_{N}$ thus obtained is effectively 
based on the assumption that the neutrino 
Yukawa couplings are large: $|\lambda_{\ell a}| \sim 1$.
The alternative possibility is to have heavy Majorana 
neutrino masses in the range of $\sim (100 \div 1000)$ GeV, 
i.e. TeV scale see-saw generation of neutrino masses.
This possibility has been recently revisited  by several authors
(see e.g. \cite{Ibarra:2010xw} and references quoted therein).

Following the discussion reported in \cite{Ibarra:2010xw}, the
low-energy neutrino physics constraints on the TeV scale type I 
see-saw scenarios are reviewed. 
In particular, it is analyzed in detail,  within three different frameworks relying on the type
I see-saw extension of the SM, the possibility of observing $|\Delta L|=2$
processes at current and future particle accelerators, LHC included, due to 
the production and decays of the heavy Majorana neutrinos.

\section{Non-unitarity effects, see-saw mechanism and $\betabeta$-decay}

Following the formalism developed in \cite{Ibarra:2010xw},
after the diagonalization of the full neutrino mass matrix in (\ref{typeI}),
the charged current (CC) and neutral current (NC)
weak interactions involving the light Majorana 
neutrinos $\chi_j$ with definite mass $m_j$ are
\begin{eqnarray}
\label{nuCC}
\mathcal{L}_{CC}^\nu 
&=& -\,\frac{g}{\sqrt{2}}\, 
\bar{\ell}\,\gamma_{\alpha}\,\nu_{\ell L}\,W^{\alpha}\;
+\; {\rm h.c.}
=\, -\,\frac{g}{\sqrt{2}}\, 
\bar{\ell}\,\gamma_{\alpha}\,
\left( (1+\eta)U \right)_{\ell i}\,\chi_{i L}\,W^{\alpha}\;
+\; {\rm h.c.}\,,\\
\label{nuNC} 
\mathcal{L}_{NC}^\nu &=& -\, \frac{g}{2 c_{w}}\,
\overline{\nu_{\ell L}}\,\gamma_{\alpha}\,
\nu_{\ell L}\,
Z^{\alpha}\;  
= -\,\frac{g}{2 c_{w}}\,
\overline{\chi_{i L}}\,\gamma_{\alpha}\,
\left (U^\dagger(1+\eta+\eta^\dagger)U\right)_{ij}\,\chi_{j L}\,
Z^{\alpha}\;.
\end{eqnarray}
The unitary matrix $U$ diagonalizes the light Majorana
mass matrix, $m_{\nu}=U^{*}\diag(m_{1},m_{2},m_{3})U^{\dagger}$.
In the basis in which the three charged lepton fields are equal to the respective
mass eigenstates, $U$ corresponds to the PMNS neutrino mixing matrix.
The hermitian matrix $\eta$, therefore, parametrizes
the deviation from unitarity of the neutrino mixing matrix due to the mixing
of the flavour neutrino fields $\nu_{\ell L}$ with the heavy RH fields in
the neutrino mass Lagrangian (\ref{typeI}). 
As shown below, $\eta$ depends on a combination of the see-saw parameters
responsible for lepton flavour violating processes.
Notice that in deriving the expressions of the CC and NC weak interactions
it is implicitly assumed that $\eta_{ij}\ll1$. Indeed,
for SM singlet fermions with masses above the EW symmetry breaking scale, i.e. bigger 
than $\sim$ 100 GeV, the combined data on
neutrino oscillation experiments and EW interaction processes provide
the upper bound \cite{Antusch:2008tz,Antusch:2006vwa}: $|\eta_{ij}|<6\times 10^{-3}$.
 
 The charged current 
and the neutral current  
interactions of the heavy  Majorana neutrino mass eigenstates
$N_j$ with $W^\pm$ and $Z^0$ formally read \cite{Ibarra:2010xw}:
\begin{eqnarray}
 \mathcal{L}_{CC}^N &=& -\,\frac{g}{2\sqrt{2}}\, 
\bar{\ell}\,\gamma_{\alpha}\,(RV)_{\ell k}(1 - \gamma_5)\,N_{k}\,W^{\alpha}\;
+\; {\rm h.c.}\,\label{NCC},\\
 \mathcal{L}_{NC}^N &=& -\frac{g}{2 c_{w}}\,
\overline{\nu_{\ell L}}\,\gamma_{\alpha}\,(RV)_{\ell k}\,N_{k L}\,Z^{\alpha}\;
+\; {\rm h.c.}\;,\label{NNC}\,
\end{eqnarray}
where the unitary matrix $V$ diagonalizes the RH neutrino mass
term in (\ref{typeI}), $M_{N}\simeq V^{*}\diag(M_{1},\ldots,M_{k})V^{\dagger}$, and
$R\equiv(M_{D} M_{N}^{-1})^{*}$.
The experimental limits on the deviations from unitarity in the neutrino mixing matrix
affect the couplings $(RV)_{\ell j}$. Indeed, in this class of models \cite{Ibarra:2010xw}: 
 $\eta=-\frac{1}{2}RR^{\dagger}$.

It is important to remark that, apart from non-unitarity effects, 
further constraints on $(RV)_{\ell j}$ and/or a particular combination
of them can be obtained at low energy considering $i)$  
the scale of the symmetric matrix $m_{\nu}=-M_{D}M_{N}^{-1}(M_{D})^{T}\simeq -R^* M_N R^\dagger\lesssim1$ eV,
resulting from the see-saw mechanism of generation of neutrino masses, and $ii)$ from the experimental 
upper bound on the effective Majorana
mass, $|(m_{\nu})_{ee}|$ \cite{bb0nudata}, which enters in the expression of the $\betabeta$-decay rate of even-even nuclei
(see e.g. \cite{BiPet87,BPP1}).
Indeed, for the heavy Majorana neutrinos  $N_k$  
with masses $M_k \sim M_R \geq 100$ GeV, barring ``accidental'' cancellations or 
extreme fine-tuning (at the level of $\sim 10^{9}$), the constraint $i)$ implies: 
\begin{equation}
 | (RV)_{\ell j}|\; \lesssim 3\times10^{-6}\left(\frac{100\; {\rm GeV}}{M_R}\right)^{1/2}\;.
\end{equation}
The estimate of $|(RV)_{\ell j}|$ thus obtained makes the heavy Majorana neutrinos $N_j$ practically 
unobservable at particle accelerators (see e.g. \cite{HanPasc09}).~In order for the CC and NC couplings of the heavy Majorana 
neutrinos $N_j$ to $W^\pm$ and $Z^0$, 
eqs (\ref{NCC}) and (\ref{NNC}), to be sufficiently large, that is $|(RV)_{\ell j}|\approx 10^{-2}$,
so that the see-saw mechanism could 
be partially or completely tested in experiments at 
the currently operating and planned future accelerators 
(LHC included), the small size of the $m_{\nu}$ matrix elements 
should be due to strong mutual compensation between the
parameters that enter in the see-saw mass relation.
In general, such cancellations arise naturally from 
symmetries in the lepton sector, 
corresponding, for instance, to the conservation 
of some additive lepton charge $\hat{L}$ 
(see, e.g. \cite{BiPet87,LeungSTP83,LWDW83,Branco:1988ex}).
In the limit in which $\hat{L}$ is exact, the RH neutrinos $\nu_{aR}$ ($a=1,\ldots,k$)
 should be Dirac particles, which is possible 
for all heavy neutrinos only if the number of the 
RH singlet neutrino fields is even: 
$k= 2q$, $q=1,2,\ldots$. If their number is odd, 
barring again ``accidental'' cancellations,
some (odd number) of the discussed 
heavy Majorana neutrinos will have
strongly suppressed couplings 
to the $W^{\pm}$ and charged leptons
and will be practically unobservable in 
the current and the future planned 
accelerator experiments. 

The resulting mass spectrum will depend
on the assumed total lepton charge $\hat{L}$. Indeed,
if $L_{i}$ ($i=e,\mu,\tau,\ldots,k$) are the ordinary lepton
charges of the fermion fields, in the basis in which the charged lepton mass matrix is diagonal,
the conserved lepton charge $\hat{L}$ takes the general form:
\begin{equation}
	\hat{L}\;=\;\sum_{i=e,\mu,\tau,\ldots,k} (-1)^{n_{i}}\,a_{i}\,L_{i}\,,\label{LC}
\end{equation}
where $n_{i}=0,1$ and $a_{i}=0,1$ for $i=e,\mu,\tau,\ldots,k$, with at least one $a_{i}\neq 0$.
In this case, it can be shown \cite{LeungSTP83} 
that  the number of massive Dirac neutrinos and massless neutrinos  in the theory is, respectively,
${\rm min}\left(n_{+}(\hat{L}),n_{-}(\hat{L})\right)$ and $\left|n_{+}(\hat{L})-n_{-}(\hat{L})\right|$, where
$n_{+}(\hat{L})$ ($n_{-}(\hat{L})$) is the number of ordinary lepton charges
in (\ref{LC}) which enter with the plus (minus) sign.

 The correct light Majorana 
neutrino mass spectrum 
can be generated by small perturbations 
that violate the corresponding symmetry $\hat{L}$.
In this case, each massive Dirac neutrino is split into
two Majorana neutrinos with close but different masses,
i.e.~the Dirac neutrinos become 
pseudo-Dirac fermions \cite{LW81,STPPD82}.
Typically, for $|(RV)_{\ell j}|\sim 10^{-3}~(10^{-4})$,
the expected splitting between the masses of the 
two heavy Majorana neutrinos forming a pseudo-Dirac pair, 
is roughly 1 (100) MeV for masses of the order of 100 (1000) GeV.
Such small mass difference will not be observable 
at LHC and planned future accelerator experiments, thus
preventing any possible experimental test at collider of the Majorana nature 
of the heavy Majorana neutrinos of the type I see-saw mechanism.

The other low energy constraint to the see-saw parameter space, strictly related to
the Majorana nature of the heavy and light Majorana neutrinos, is provided by
$\betabeta$-decay experiments \cite{bb0nudata}. In this case, in addition 
to the standard contribution due to the light Majorana 
neutrino exchange, the  $\betabeta$-decay effective 
Majorana mass $|(m_{\nu})_{ee}|$ 
receives a contribution from the exchange of the heavy Majorana 
neutrinos $N_k$:
\begin{equation}
|(m_{\nu})_{ee}| \cong 
\left |\sum_{i}(U_\PMNS)^2_{ei}\, m_i 
- \sum_k\, F(A,M_k)\, (RV)^2_{e k}\,M_k \right |\,,
\label{mee1}
\end{equation}
%
where $F(A,M_k)$ is a known real (positive) function
of the atomic number $A$ of the decaying nucleus
and of the mass $M_k$ of $N_k$ \cite{HPR83,JV83,HaxStev84}.
For the range $M_k \sim (100 \div1000)$ GeV of interest, 
one has with good approximation (see, e.g. \cite{JV83,HaxStev84,Hirsch01,Blennow:2010th})
$F(A,M_k) \simeq (M_a/M_k)^2 f(A)$, with $M_{a}\simeq 0.9$ GeV and 
$f(A)\simeq\mathcal{O}(10^{-2})$.

\section{The case of a broken lepton charge $\hat{L}$}

It is shown in the following that, within the type I see-saw scenario
of generation of light neutrino masses, if 
$m_{\nu}\neq 0$ arises as result of the breaking of a global symmetry
corresponding to some conserved lepton charge $\hat{L}$,
the Majorana nature of the heavy Majorana neutrinos could be hardly tested at collider physics.
As stated before, the reason is that if the heavy Majorana fields 
involved in the see-saw mechanism of LH neutrino mass generation,
have couplings to charged leptons and weak gauge bosons sufficiently large 
to produce and observe them at collider, they
would behave effectively as a pseudo-Dirac fermion and all lepton number 
violating processes, e.g.~same sign di-muon production in proton-proton
collisions, would be strongly suppressed.

Consider for simplicity the case  of three RH neutrinos
$\nu_{aR}$ ($a=1,2,3$) and a conserved lepton charged $\hat{L}=L_{e}+L_{\mu}+L_{\tau}+L_{3}-L_{1}$ 
(see e.g. \cite{Ibarra:2010xw}).
A possible choice of the Dirac and Majorana mass matrix which preserves $\hat{L}$ is:
\begin{equation}
	M_D = \left(
     \begin{array}{ccc} 
  0 & 0 & m^{D}_{e3} \\
     0   & 0 & m^{D}_{\mu3}\\
     0   & 0 & m^{D}_{\tau 3}
\end{array}\right)\,,\;\;\;\;\;\;\;\;
M_N = \left(
     \begin{array}{ccc} 
   M_{11} & 0 & 0 \\
     0   & 0 & M_{23}\\
     0   & M_{23} & 0
\end{array}\right)\,,\label{see-saw masses}
\end{equation}
where, without loss of generality, all the parameters in $M_{D}$ and $M_{N}$
are taken real and positive.
As a consequence of the simplifying 
choice made, $m^{D}_{\ell 1} = 0$, $l=e,\mu,\tau$,
the Majorana field $N_{1} = (\nu_{1R} + \nu^{C}_{1L})/\sqrt{2}$ is decoupled
and does not take part in the see-saw mechanism of generation of LH neutrino masses.
In this case $n_{+}(\hat{L})=4$ and $n_{-}(\hat{L})=1$ and
the spectrum of the theory contains 3 massless neutrinos and 1 massive Dirac fermion.
The latter is given by the combination $N_{D}=(N_{2}+N_{3})/\sqrt{2}$ and
has a mass $M = M_{23} > 0$, 
where the two heavy Majorana neutrinos 
$N_{2}$ and $N_{3}$ have the same mass 
$M_{2} = M_{3} = M_{23}$  
and satisfy the Majorana conditions   
$C \overline{N_{k}}^T =\rho_k N_k$, 
$k=2,3$, with $\rho_2 = -1$ and $\rho_3 = +1$.

In order to have a light neutrino mass spectrum compatible with 
the observations, it is necessary to softly break $\hat{L}$.~Taking 
$(M_{D})_{\ell 2}\equiv m^{D}_{\ell 2}\neq 0$, it is easy to show  \cite{Ibarra:2010xw} that
two of the light neutrinos acquire a finite mass, $0<m_{2}<m_{3}$,
while the Dirac fermion $N_{D}$ is split into two different Majorana fields
$N_{2}$ and $N_{3}$ with $\Delta M\equiv M_{3}-M_{2}=m_{3}-m_{2}=2A/M_{23}$,
where $A$ depends explicitly on the lepton charge breaking parameters $m^{D}_{\ell 2}$:
$A =  m^{D}_{e2}m^{D}_{e3}+m^{D}_{\mu 2} m^{D}_{\mu 3}+ m^{D}_{\tau 2} m^{D}_{\tau 3}$.
The light neutrino mass scale, $m_{3}-m_{2}\lesssim 1$ eV, imposes
strong constraints on the symmetry breaking parameters.
More explicitly, for $M_{23}\approx 100$ GeV one has:
$|m^{D}_{\ell 2} m^{D}_{\ell^{\prime} 3}| \ltap 10^{-7} {\rm GeV}^{2}$ ($\ell,\ell^{\prime}=e,\mu,\tau$).
Because of the small mass splitting  $\Delta M$, the heavy Majorana neutrinos $N_{2,3}$  
effectively behave in the collider physics as a pseudo-Dirac fermion $N_{PD}=(N_{2}+N_{3})/\sqrt{2}$. 
 
 Consider now the process of same sign di-muon 
production in proton-proton collisions, assuming that 
one of the muons, say $\mu^-$, is produced together 
with real or virtual $N_{2,3}$ in the decay of 
a virtual $W^-$, while the second $\mu^-$ 
originates from  the decay $N_{2,3} \rightarrow W^+ \,\mu^-$,
with virtual or real $W^+$ which decays further
into, e.g.~two hadronic jets.~In order to collect a sufficiently
large number of events, it is necessary to 
have sizable couplings of the heavy Majorana neutrinos to the muon field.
The latter can be easily computed \cite{Ibarra:2010xw}: $(RV)_{\mu 2}=m^{D}_{\mu 3}\cos\theta /M_{23}$
and $(RV)_{\mu 3}=m^{D}_{\mu 3} \sin\theta/M_{23}$, with $\tan^{2}\theta=M_{2}/M_{3}$. In the present
scenario $\theta\simeq\pi/4$ because $N_{2}$ and $N_{3}$ form a pseudo-Dirac fermion. If, for instance,
$M_{23}\approx100$ GeV and $|(RV)|_{\mu 2,3}\approx 10^{-2}$, one obtains
$m^{D}_{\mu 3}\approx 1$ GeV and $\hat{L}$ charge breaking parameter $m^{D}_{\mu 2}\approx 100$ eV,
due to the constraint which arises from the small neutrino masse scale (see the discussion above).
The relevant part for the amplitude of the process under discussion is \cite{Ibarra:2010xw}:
\begin{equation}
A(p\;p\;\to\;\mu^{-}\;\mu^{-}\;2{\rm jets}\;X) \propto \frac{(m^D_{\mu 3})^2}{M^2_{23}}\, \frac{M_2M_3}{M_3 + M_2}\,
\frac{M^2_3 - M^2_2 - i( \Gamma_3 M_3 - \Gamma_2 M_2)}
{(p^2 - M^2_3 + i\Gamma_3 M_3)(p^2 - M^2_2 + i\Gamma_2 M_2)}\,.
\label{ppmumu2}
\end{equation}
%
Taking into account that $M_2M_3 \simeq M^2_{23}$  and 
$\Gamma_{2(3)} \propto \sum_{\ell} |(RV)_{\ell 2,3}|^{2} G_F\, M^3_{2(3)}$,
it is easy to show that the amplitude for this lepton number violating process is proportional to the factor
 $\Delta M/M_{23}$. As discussed previously, in the pseudo-Dirac scenario of interest, which is consequence of an almost conserved
 lepton charge present in the theory, the heavy Majorana neutrino mass splitting is strongly bounded
 by the light neutrino mass scale, $\Delta M\lesssim1$ eV,
 thus making the same sign di-muon production at LHC unobservable.

\section{$\hat{L}$ is not conserved but $m_{\nu} = 0$ at leading order} 

A simple realization of this scenario is obtained from the mass textures reported
in eq.~(\ref{see-saw masses}), in which the Majorana mass matrix, $M_{N}$, is conveniently modified
with the $\hat{L}$-breaking mass term: $(M_{N})_{33}=M_{33}>0$.
In this case, even if lepton number is not conserved, at leading order one has
$m_{\nu}=M_D\,M^{-1}_N\,M^T_D = 0$ and, therefore, the tree-level constraints on the see-saw parameter 
space from the light neutrino mass scale do not apply here:
light neutrino masses, compatible with neutrino data, can be generated at higher order (one or two loops). 
This in turn could lead to sufficiently large
$N_{2,3}$ production  rates at LHC 
to make the observation of 
the two heavy Majorana neutrinos 
possible. Indeed, in such scenario
$\Delta M=M_{3}-M_{2}=M_{33}$ and, in principle, one can
observe the same sign di-muon production at LHC for a sufficiently large
mass splitting (see eq.~(\ref{ppmumu2})).
Anyway, it should be clear from the preceding discussion 
that all $|\Delta L| = 2$ Majorana type effects
should vanish in the limit of  $M_{33} = 0$. 

As discussed before, in such class of models,
the heavy Majorana neutrinos might provide a sizable contribution to the
\betabeta-decay rate. Indeed, from eq.~(\ref{mee1}),  neglecting the standard contribution due to
the tree-level exchange of light Majorana neutrino mass eigenstates, the effective Majorana
mass $|(m_{\nu})_{ee}|$ results proportional to the mass splitting $\Delta M$ \cite{Ibarra:2010xw}:
\begin{eqnarray}
|(m_{\nu})_{ee}| &\cong& \left | \frac{(m^D_{e3})^2}{M^2_{23}}\,
f(A)\, M_a^2\, \frac{M_3-M_2}{M_2\,M_3}\right |
\cong \left |\frac{(m^D_{e3})^2}{M^2_{23}}\,f(A)\,
\frac{M^2_a}{M^2_{23}}\,M_{33} \right |\,.
\label{mee3}
\end{eqnarray}
If  $(m^D_{e3})^2/M^2_{23}$ will be determined from 
an independent measurement, the bound on  $|(m_{\nu})_{ee}|$ 
will lead to a strong constraint on $M_{33}/M_{23}$. 
Taking e.g.~$|(m_{\nu})_{ee}| \ltap 1$ eV, $f(A) = 0.078$ (corresponding 
to $^{76}$Ge) and the maximal value of 
$(m^D_{e3})^2/M^2_{23}$ allowed by the non-unitarity constraints (see \cite{Antusch:2008tz}), 
$8\times 10^{-3}$, one finds: 
$M_{33}\ltap 1.8 \times 10^{-5}\, M_{23} (M_{23}/M_a)$.~For $M_{23}=100$ GeV this implies  
$M_{33}\ltap 2\times 10^{-3} M_{23} 
\cong 0.2~{\rm GeV} \ll M_{23}$.~Such a small mass difference would render the 
Majorana-type effects associated with $N_{2,3}$
hardly observable.~If, however, 
$|(m^D_{e3}/M_{23})^2| \ltap 1.6\times 10^{-6}$,
$M_{33}\ltap M_{23}$ and a signature of 
the process $p\,p\rightarrow \mu^- \, \mu^- \,2~{\rm jets} \, X$,
generated by the production and decay of 
real or virtual $N_{2,3}$, is detectable at LHC.

\section{The extreme fine-tuning case} 

Consider now the type I see-saw Lagrangian in which the
RH neutrino masses are given between ($100\div 1000$) GeV and
the see-saw mechanism provides the observed LH neutrino mass scale and mixing
without assuming any softly broken conserved lepton charge in the theory.
This possibility requires, in general, a huge tuning of parameters. Namely,
demanding $(M_{D})_{ij}\sim{\cal O}(1\,{\rm GeV})$
and $M_j \sim{\cal O}(100\,{\rm GeV})$ requires a tuning of one
part in $10^9$ in order to produce a neutrino mass 
$m_i \sim{\cal O}(10^{-2}\,{\rm eV})$ \cite{Ibarra:2010xw}.

Unlike the case discussed earlier, now the see-saw parameter space 
is constrained by both the small light neutrino mass scale, $m_{\nu}\lesssim 1$ eV, and the upper bound on the
effective Majorana mass in \betabeta-decay experiments. Consider, for instance,
the two heavy Majorana neutrinos mass eigenstates $N_{2}$ and $N_{3}$ with masses
$M_{2}>0$ and $M_{3}\equiv M_{2}(1+z)$, $z>0$, and satisfying standard Majorana conditions,
$C\overline{N}_{2,3}^{T}=N_{2,3}$. For large neutrino Yukawa couplings,
the CC and NC weak interactions of $N_{2}$ and $N_{3}$ with charged leptons and 
gauge bosons, eqs~(\ref{NCC}) and (\ref{NNC}), satisfy, with a good approximation, 
 the relation \cite{Ibarra:2010xw}: $(RV)_{\ell 3}\simeq i(RV)_{\ell 2}/\sqrt{1+z}$ ($\ell=e,\mu,\tau$),
 with $|(RV)_{\mu 2}|\approx |(RV)_{e2}|$. 
 From eq.~(\ref{mee1}), neglecting again the contribution due to the exchange of the light Majorana
 neutrino mass eigenstates, the effective Majorana mass is given by
 \begin{equation}
	|(m_{\nu})_{ee}| \cong 2 z\,\left| \left(RV\right)_{e2} \right|^{2}\, \frac{M_{a}^{2}}{M_{2}}\,f(A)\,.\label{mee3}
\end{equation} 
Assuming production and detection of the RH Majorana fields at collider, i.e. $|(RV)_{\ell 2}|\approx 10^{-2}$ ($\ell=e,\mu$), one has that
$|(m_{\nu})_{ee}|\lesssim0.2$ eV for $z\lesssim10^{-3}\,(10^{-2})$ and $M_{1}\approx 100\, (1000)$ GeV.
Therefore, the non-observation 
of the $(\beta\beta)_{0\nu}$-decay requires in this scenario a 
degeneracy in the right-handed neutrino masses of at least one per cent.
As discussed above, the cross section for same sign di-muon production
in proton-proton collisions is proportional to the mass difference of the 
right-handed neutrinos. Thus, even in this extremely fined-tuned scenario, 
the Majorana nature of the right-handed neutrinos will be difficult to probe
at colliders.

\section{Conclusions}

The see-saw mechanism provides
a natural explanation for the smallness of neutrino masses.
In such scenario, a new mass scale is introduced in the theory, which is linked to
the light neutrino mass scale and 
can in principle be accessible to present and forthcoming particle 
physics accelerators, LHC included.
In its simplest formulation, the
type I see-saw scenario, the Standard Model (SM) is 
extended with at least two 
``heavy'' Majorana neutrinos,
which are singlets of $SU(2)_{L}\times U(1)_{Y}$ and
are coupled to the SM lepton and Higgs doublets.
It is shown that, if the type I see-saw scenario provides 
the correct mechanism of generation of light neutrino masses and
the Majorana fields $N_{j}$ have masses 
in the range $M_j \sim (100 \div 1000)$ GeV,
the physical effects associated with 
the Majorana nature of these heavy 
neutrinos $N_j$, 
are so small that they
are unlikely to be observable
in the currently operating and 
future planned accelerator experiments 
(including LHC). This is a consequence
of the existence of very strong constraints 
on the parameters and couplings, responsible for the 
corresponding $|\Delta L| = 2$ processes in which 
$N_j$ are involved, and/or
on the couplings of $N_j$ 
to the weak $W^{\pm}$ 
and  $Z^0$ bosons. 
The strongest constraints are provided by
the experimental upper limit on $i)$ the light 
neutrino mass matrix, $|(m_{\nu})_{ll'}| \lesssim 1$ eV, $l,l'=e,\mu,\tau$
and $ii)$ the effective Majorana mass $|(m_{\nu})_{ee}|$
obtained in the $\betabeta$-decay 
experiments. The latter, is sensitive to the 
heavy Majorana neutrino mass scale and 
splitting(s). 

As consequence, it is shown that 
charged and neutral current interactions 
of the heavy Majorana fields $N_j$ with the Standard Model charged 
leptons and neutrinos are extremely suppressed, unless the heavy Majorana neutrinos form
a pseudo-Dirac pair.
Therefore, the Majorana nature of the heavy
neutrinos will not be detected in collider experiments: either the
production cross section is highly suppressed or the heavy neutrinos
behave to a high level of precision as Dirac fermions.

The suppression of the $|\Delta L|=2$ processes can be avoided only 
if there exist  additional TeV scale interaction terms in the Lagrangian between 
the heavy Majorana neutrinos and the Standard Model particles. 
If this is the case, the production cross section of heavy 
neutrinos will not necessarily be suppressed, while 
their charged and neutral current interactions 
with the Standard Model charged leptons and neutrinos can still 
be tiny.

\section*{Acknowledgements}
Emiliano Molinaro would like to thank the organizers of the conference
DISCRETE 2010, Symposium on Prospects in the Physics of Discrete Symmetries,
for the opportunity of presenting the topics discussed in these Proceedings.
The work of Emiliano Molinaro is supported by the Funda\c{c}\~{a}o para a Ci\^{e}ncia e a
Tecnologia (FCT, Portugal) through the project
PTDC/FIS/098188/2008 and CFTP-FCT Unit 777 which are partially funded
through POCTI (FEDER).


\section*{References}

\begin{thebibliography}{9}
\bibitem{exper}
B.T. Cleveland  {\it et al.}, Astrophys.\
J. \textbf{496}, 505 (1998);
Y. Fukuda {\it et al.} [Kamiokande Collaboration],
Phys.\ Rev.\ Lett. \textbf{77}, 1683 (1996);
J.N. Abdurashitov {\it et al.}, Phys.\ Rev.\ C \textbf{80},
015807 (2009);
P. Anselmann {\it et al.}, 
Phys.\ Lett.\ B \textbf{285} (1992) 376;
W. Hampel {\it et al.}, 
Phys.\ Lett.\ B \textbf{447}  (1999) 127;
M. Altmann {\it et al.}, Phys.\ Lett.\ 
B \textbf{616}  (2005) 174;
 S.~Fukuda {\it et al.} [Super-Kamiokande Collaboration],
Phys. Lett. {\bf B539} (2002) 179;
 Q.R. Ahmad {\it et al.} [SNO Collaboration], 
Phys.\ Rev.\ Lett. \textbf{87} (2001) 071301 and
\textbf{89} (2002) 011301;
Y.~Fukuda {\it et al.}  [Super-Kamiokande Collaboration],
Phys.\ Rev.\ Lett.\  {\bf 81} (1998) 1562;
Y. Ashie {\it et al.} [Super-Kamiokande Collaboration], 
Phys. Rev. Lett. {\bf 93} (2004) 101801;
K.~Eguchi {\it et al.}  [KamLAND Collaboration],
Phys.\ Rev.\ Lett.\  {\bf 90} (2003) 021802; 
T. Araki {\it et al.},
Phys.\ Rev.\ Lett. \textbf{94} (2005) 081801;
C. Arpesella {\it et al.}, 
Phys. Lett. B {\bf 658} (2008) 101; 
Phys. Rev. Lett. {\bf 101} (2008) 091302;
M.~H.~Ahn {\it et al.}  [K2K Collaboration],
Phys.\ Rev. \ D {\bf 74} (2006) 072003;
D.G. Michael {\it et al.} [MINOS Collaboration], 
Phys.\ Rev.\ Lett. \textbf{97} (2006) 191801; 
P. Adamson {\it et al.},
Phys.\ Rev.\ Lett. \textbf{101} (2008) 131802.

\bibitem{BPont57} B. Pontecorvo, 
                  Zh. Eksp. Teor. Fiz. (JETP) {\bf 33} (1957) 549 
                and {\bf 34} (1958) 247.

\bibitem{MNS62} Z. Maki, M. Nakagawa and S. Sakata, 
Prog. Theor. Phys. {\bf 28} (1962) 870.

\bibitem{Znu}  
  C.~Amsler {\it et al.}  [Particle Data Group],
  Phys.\ Lett.\  B {\bf 667} (2008) 1.


\bibitem{BPont67}
  B.~Pontecorvo,
  Sov.\ Phys.\ JETP {\bf 26} (1968) 984.
  [Zh.\ Eksp.\ Teor.\ Fiz.\  {\bf 53}, 1717 (1967)].

\bibitem{seesaw}  P. Minkowski,
Phys.\ Lett.\  B {\bf  67} (1977) 421;
M. Gell-Mann, P. Ramond and R. Slansky, 
{\em Proceedings of the Supergravity Stony Brook Workshop}, 
New York 1979,  eds. P. Van Nieuwenhuizen and D. Freedman;
T. Yanagida,  
{\em Proceedinds of the Workshop on Unified Theories and Baryon Number in theUniverse},  Tsukuba, Japan 1979, ed.s A. Sawada and A. Sugamoto;
 R. N. Mohapatra and G. Senjanovic, Phys. Rev. Lett. {\bf 44} (1980) 912.
  
\bibitem{Ibarra:2010xw}
  A.~Ibarra, E.~Molinaro and S.~T.~Petcov,
  JHEP {\bf 1009} (2010) 108
  [arXiv:1007.2378 [hep-ph]].
  
\bibitem{Antusch:2008tz}
  S.~Antusch, J.~P.~Baumann and E.~Fernandez-Martinez,
  Nucl.\ Phys.\  B {\bf 810} (2009) 369.

\bibitem{Antusch:2006vwa}
  S.~Antusch {\it et al.},
  JHEP {\bf 0610} (2006) 084.
  
  
\bibitem{bb0nudata} See, e.g. C. Aalseth {\it et al.}, 
arXiv:hep-ph/0412300;
C.E. Aalseth {\it et al.} [IGEX Collaboration],
Phys. Atomic Nuclei {\bf 63} (2000) 1225;
H.~V.~Klapdor-Kleingrothaus {\it et al.},  
Phys. Lett. B {\bf 586} (2004) 198;
A.~S.~Barabash  [NEMO Collaboration],
arXiv:0807.2336 [nucl-ex];
C.~Arnaboldi {\it et al.}  [CUORICINO Collaboration],
Phys.\ Rev.\  C {\bf 78} (2008) 035502.


  
\bibitem{BiPet87} S.~M.~Bilenky and S.~T.~Petcov,
Rev.\ Mod.\ Phys.\  {\bf 59} (1987) 671.


  
\bibitem{BPP1}  S.M. Bilenky, S. Pascoli and S.T. Petcov,
Phys. Rev. D {\bf 64} (2001) 053010;
S.T. Petcov, Phys. Scripta T {\bf 121} (2005) 94;
S. Pascoli and S. T. Petcov,
Phys. Rev. D {\bf 77} (2008) 113003. 

 \bibitem{HanPasc09} A. Atre {\it et al.},
JHEP {\bf 0905} (2009) 030.


\bibitem{LeungSTP83} C.N. Leung and S.T. Petcov, 
Phys. Lett. B {\bf 125} (1983) 461.

\bibitem{LWDW83} 
D.~Wyler and L.~Wolfenstein,
Nucl. Phys.  B {\bf 218} (1983) 205.

\bibitem{Branco:1988ex}
  G.~C.~Branco, W.~Grimus and L.~Lavoura,
  Nucl.\ Phys.\  B {\bf 312} (1989) 492.
  
\bibitem{LW81} L. Wolfenstein, Nucl. Phys. B {\bf 186} (1981) 147.

\bibitem{STPPD82} S.T. Petcov, Phys. Lett. B {\bf 110} (1982) 245.

\bibitem{HPR83} A. Halprin, S.T. Petcov and S.P. Rosen, 
Phys. Lett. B {\bf 125} (1983) 335.

\bibitem{JV83} J. Vergados, Nucl. Phys. B {\bf 218} (1983) 109.

\bibitem{HaxStev84} W.C. Haxton and J. Stephenson,
Prog. Part. Nucl. Phys. {\bf 12} (1984) 409.

\bibitem{Hirsch01} H. Paes {\it et al.},
Phys. Lett. B {\bf 498} (2001) 35.


\bibitem{Blennow:2010th}
  M.~Blennow et {\it al.},
  JHEP {\bf 1007} (2010) 096
  [arXiv:1005.3240 [hep-ph]].

\end{thebibliography}
\end{document}